\documentclass[prd,twocolumn,showpacs,nofootinbib,preprintnumbers,floatfix]{revtex4}  
\usepackage[plainpages=false, colorlinks=true, anchorcolor=blue, linkcolor=blue, citecolor=blue, bookmarks=false]{hyperref}
\usepackage{graphicx}  
\usepackage{dcolumn}   
\usepackage{bm}        
\usepackage{amssymb}   
\usepackage{natbib}
\usepackage{amsmath}   
\usepackage{longtable}  
\usepackage{float} \usepackage[utf8]{inputenc}
\usepackage{tasks}

\begin{document}

\newcommand{\apjl}{Astrophys. J. Lett.}
\newcommand{\apjs}{Astrophys. J. Suppl. Ser.}
\newcommand{\aap}{Astron. \& Astrophys.}
\newcommand{\rthis}[1]{\textcolor{black}{#1}}
\newcommand{\aj}{Astron. J.}
\newcommand{\pasp}{PASP}
\newcommand{\araa}{Ann. Rev. Astron. Astrophys. } 
\newcommand{\aapr}{Astronomy and Astrophysics Review}
\newcommand{\ssr}{Space Science Reviews}
\newcommand{\mnras}{Mon. Not. R. Astron. Soc.}
\newcommand{\apss} {Astrophys. and Space Science}
\newcommand{\jcap}{JCAP}
\newcommand{\na}{New Astronomy}
\newcommand{\pasj}{PASJ}
\newcommand{\pasa}{Pub. Astro. Soc. Aust.}
\newcommand{\physrep}{Physics Reports}

\title{A test of the  standard dark matter density evolution law using galaxy clusters  and cosmic chronometers}

\author{Kamal Bora$^{1}$}\email{ph18resch11003@iith.ac.in}

\author{R. F. L. Holanda$^{2,3,4}$}\email{holandarfl@fisica.ufrn.br}

\author{Shantanu Desai$^{1}$}\email{shntn05@gmail.com}

\author{S. H. Pereira$^{5}$}\email{s.pereira@unesp.br}

\affiliation{$^1$ Department of Physics, Indian Institute of Technology, Hyderabad, Kandi, Telangana-502285, India }

\affiliation{$^2$Departamento de F\'{\i}sica, Universidade Federal do Rio Grande do Norte,Natal - Rio Grande do Norte, 59072-970, Brasil}

\affiliation{$^3$Departamento de F\'{\i}sica, Universidade Federal de Campina Grande, 58429-900, Campina Grande - PB, Brasil}

\affiliation{$^4$Departamento de F\'{\i}sica, Universidade Federal de Sergipe, 49100-000, Aracaju - SE, Brazil}

\affiliation{$^5$ Departamento de F\'isica, Faculdade de Engenharia de Guaratinguet\'a, Universidade Estadual Paulista (UNESP), 12516-410, Guaratinguet\'a - SP, Brazil}

\begin{abstract}
In this paper, we implement a  test of the standard law for  the dark matter density evolution as a function of redshift.  For this purpose, only a flat universe and the  validity  of the FRW metric  are assumed.  A deformed dark matter density evolution  law is considered, given by  $\rho_c(z) \propto (1+z)^{3+\epsilon}$, and constraints on $\epsilon$ are obtained by combining the  galaxy cluster gas mass fractions with cosmic chronometers measurements. We find that $\epsilon =0$ within 2$\sigma$ c.l., in full agreement with other recent analyses. 
\end{abstract}

\maketitle

\section{Introduction}

The standard model of Cosmology~\cite{Ratra08}, or the concordance model,  consists of 5\% of ordinary baryonic matter, formed by all kinds of matter and radiation that  we can see directly and/or detect by several methods, about 25\% consisting of non-interacting cold dark matter (CDM), which can be measured just by its gravitational interactions and finally the remaining 70\% of dark energy (DE), consisting of a smooth fluid whose net stress energy tensor  leads to  gravitational   repulsive, and can attributed to a constant vaccuum energy density $\Lambda$~\cite{Weinberg89}. Such a three component model  is in good agreement with recent Planck CMB observations~\cite{Planck18}. Although this is a very robust model at the background level, there are also important peculiarities of the standard model that deserve further evaluation, such as the composition and failure to detect CDM in the laboratory, the core-cusp problem, Radial acceleration relation in galaxies, the missing satellite problems at small scales, problems with Big-Bang Nucleosynthesis related to Lithium-7 abundance, the theoretical discrepancy related to cosmological constant, the measure of present Hubble constant and $\sigma_8$ tension, low-$l$ anomalies in the Cosmic Microwave Background,  the cosmic coincidence problem~\cite{Caldwell2009,Weinberg2013,Bullock,Copi,Merritt,Fields,Desai18,Divalentino21,Peebles21,Mohanty,Griest}. The cosmic coincidence problem refers to why we live at special moment of evolution. This special moment is when the  CDM and DE densities are exactly of  the same order of magnitude. An uptodate review of all the  tensions within the  standard $\Lambda$CDM model of Cosmology can be found in ~\citet{Periv}. Therefore, for all these reasons, several alternative models to the $\Lambda$CDM have been proposed recently to deal with some of these issues~\cite{alternatives,Banik,Melia}.

Specifically considering the cosmic coincidence problem~\cite{Griest}, the introduction of an interaction term between the dark sectors of the universe has been shown to be an interesting alternative to solve or  alleviate the problem, as extensively discussed in ~\cite{Wang2016} and more recently in~\cite{Jesus:2020tby}.
The possibility of a coupling between dark matter and dark energy has been recently investigated  in a number of works~\cite{Ozer1985,amendola2000,barrow06,caldera09,alcaniz12}.  In \cite{marttens20b}, it was shown that a model of dynamical dark energy and  interacting
DE-DM approaches become indistinguishable at the background and linear perturbation level. There are also studies involving the dark sector interaction using both model-dependent and model-independent approaches, some of which are  also related to the Hubble tension~\cite{marttens19,marttens20a,carneiro19,clark21,supriya20,yang19,valentino17,vattis19,bonilla21}. 

 As a direct consequence of the interacting DE-DM models, an investigation on possible departures for  the evolution of the dark matter density with respect to the usual $a^{-3}$ scaling was done in~\cite{Holanda:2019sod} by combining Type Ia supernovae (SNe Ia) observations and gas mass fraction (GMF) measurements in galaxy clusters. This data could neither confirm nor rule out such an interaction. More recently, a new method to explore a possible departure from the standard time evolution law for the dark matter density was proposed~\cite{boraDM}. By using a deformed evolution law proportional to $(1+z)^{3+\epsilon}$, where $\epsilon \neq 0$ represents the violation of standard model, the aforementioned work has  shown that $\epsilon$ is consistent with zero within 1$\sigma$ c.l.
 The dataset used for this analysis consisted of  Strong Gravitational Lensing (SGL) data and gas mass fraction measurements of galaxy clusters. The lens profiles in the SGL systems can be segregated into  three sub-samples, based on their stellar velocity dispersion values. The combined analyses with GMF data showed a negligible departure of the standard law ($\epsilon = -0.088\pm 0.11$).
 
 In this paper, we propose and implement another test to probe a possible evolution  of the dark matter density law ($\rho_c(z) \propto (1+z)^3$) using  cosmic chronometers along with X-ray gas mass fraction data from~\citet{Mantz:2014xba}, to complement our previous works~\cite{Holanda:2019sod,boraDM}. 
 This search for a possible departure from the  standard evolution law is carried out  by adding an ad-hoc term ($\epsilon$), which is a function of the cosmic scale factor i.e. $\epsilon(a)$, such as $\rho_c(z) \propto (1+z)^{3+\epsilon}$ \cite{wang05,alcaniz05}. Such an evolution could arise  from a non-gravitational interaction between the dark sectors. In the first part of our analyses, we posit  the standard  dark matter density  evolution law ($\epsilon=0$) and verify if the gas depletion factor ($\gamma(z)$), i.e. the ratio by which the gas mass fraction of galaxy clusters is depleted with respect to the  universal  mean of baryon fraction, is constant with redshift  considering $\gamma(z)=\gamma_0(1+\gamma_1 z)$ and the data set. This is the simplest extension one can consider.   Since we obtain $\gamma_1=0$ within 1$\sigma$ c.l., in the second part of this work we propose a  test to probe a possible evolution  of dark matter density law. Our main conclusion  is that $\epsilon =0$ within 2$\sigma$ c.l., in full agreement with other recent  results.

This manuscript is organized as follows. In Section~\ref{data}, we briefly present the data sample used for this analysis. Section~\ref{methodology} describes the methodology followed in this work. Our analysis and results are discussed in Section~\ref{sec:analysis}. Our conclusions are presented in Section~\ref{sec:conclusions}.

\section{Data Sample}
\label{data}

\subsection{Gas Mass Fraction}

The Chandra X-ray sample used for this analysis, consists of 40 galaxy clusters~\cite{Mantz:2014xba}, spanning the redshift range $0.078 \leq z \leq 1.063$ identified through a comprehensive search of the Chandra archive for hot ($kT \geq  5$ keV), massive and morphologically relaxed systems. The choice of relaxed systems minimize the systematic biases in the hydrostatic masses. The gas mass fraction $f_{gas}$ is calculated in the spherical shell $0.8 \leq r/r_{2500} \leq 1.2$ rather than the cumulative fraction integrated over all radii ($< r_{2500}$). It excludes the innermost regions of the clusters, since these regions contain  complex structures  even in the most relaxed clusters. A more detailed discussion about the data can be found in~\cite{Mantz:2014xba}\footnote{While this work was in progress,  this dataset has been recently augmented with new measurements~\cite{Mantz21}. In this work we use the older data for our analysis. The new dataset will be analyzed in our future works.}
The gas mass fraction data comprises of five clusters at $z<0.16$. These low redshift data have been used to obtain  a  bound on the quantity: $\left(\frac{h^{3/2}\Omega_{b0}}{\Omega_{c0}+\Omega_{b0}}\right) = 0.089 \pm 0.012$, where $h$, $\Omega_{c0}$, and $\Omega_{b0}$ are the Hubble parameter, dark matter density, and baryon density at $z=0$ in units of critical density, respectively. Using $h= 0.732 \pm 0.013$ from the latest SH0ES results~\cite{Riess:2020fzl} and $100 \Omega_{b0}h^2 = 2.235 \pm 0.033$ from Big Bang Nucleosynthesis \cite{Cook2018} along with the aforementioned constraint, we get: $\rho_{b0}= (4.20 \pm 0.22) \times 10^{-31} g/cm^{3}$, $\rho_{c0}= (25.34 \pm 4.35) \times10^{-31}  g/cm^3$ and $\rho_{m0}=29.53 \pm 4.34(\times 10^{-31}  g/cm^3)$. We shall use these values in our analysis outlined in  Sec~\ref{sec:analysis}. Furthermore, we exclude the five gas mass fraction data at $z<0.16$, and thereby use only the remaining 35 $f_{gas}$ data for the analysis.

\subsection{Cosmic Chronometers}

Cosmic chronometers (CC) are one of the most widely used probes in Cosmology for deducing the observational value of the Hubble parameter at different redshifts (See ~\cite{Haveesh} and references therein).
According to \citet{jimenez}, if passively evolving galaxies at different redshifts are considered, then computing the age difference of the galaxies yields the Hubble parameter based on the following equation:
\begin{equation}
    H(z) = - \frac{1}{1+z} \frac{dz}{dt}.
\label{eq:chrono}
\end{equation}
The derivative term $dz/dt$  in Eq.~\ref{eq:chrono} is obtained with respect to the cosmic time. This method is agnostic with respect  to the choice of the cosmological model used,  given that the only assumption for the CC is the stellar population model.

We use 31 cosmic chronometer $H(z)$ data from \citet{li19} in the redshift range $0.07 \leqslant z \leqslant 1.965 $ in order to derive the angular diameter distance to the cluster (see next section).

\section{Methodology}
\label{methodology}
In a first part of our analyses, we shall  check if the gas depletion factor ($\gamma(z)$)  is constant with redshift, by considering a linear evolution law: $\gamma(z)=\gamma_0(1+\gamma_1z)$. In the  second part of this work, using these results for  $\gamma$, we  implement a test of  a possible evolution  of the dark matter density law  using   galaxy clusters  and cosmic chronometers.

\subsection{Gas depletion factor}
\label{depletion factor}

The matter content of galaxy clusters should approximately match the matter content of the Universe as a whole~\cite{borgani11}. 
Therefore the gas mass fraction which is defined as the ratio of baryonic matter density $\Omega_b$ to the total matter density $\Omega_m$ of the universe should be a constant over the cosmic time for massive and relaxed clusters. Therefore, this quantity can be used as a cosmological probe~\cite{Sasaki,Pen97}. The gas mass fraction  is given by~\cite{Allen2007},
\begin{equation}
\label{fgas}
f_{gas}(z) = K(z)A(z)\gamma(z)  \left[\frac{\Omega_b(z)}{\Omega_m(z)}\right] \left(\frac{D_A^*}{D_A}\right)^{3/2}.\
\end{equation}
Here, $K(z)$ is the calibration constant, which accounts for any bias in gas mass due to bulk motion and non-thermal pressure in cluster gas~\cite{Allen2007,Mantz:2014xba,Planelles2013,Battaglia2013}. This value has been obtained self-consistently from the data by comparing with clusters having weak lensing masses from Weighing the Giants sample~\cite{app}, and is equal to $0.9 \pm 0.09$.
$A(z)$ represents the angular correction factor representing the change in angle subtended at $r_{2500}$ as the cosmology is varied~\cite{Allen2007}. Since this factor is close to unity~\cite{Allen2007}, we do not incorporate it in our analysis.
and $D_A$ is the angular diameter distance to each cluster. The  asterisk in Eq.~\ref{fgas} denotes the corresponding quantity when a flat $\Lambda$CDM fiducial cosmology ($H_0$ = 70 km/sec/Mpc and $\Omega_m$ = 0.3) is assumed. The most relevant quantity for this work is the gas depletion factor $\gamma(z)$, which is a measure of  how much baryonic gas is depleted compared to  the cosmic mean. Hence from Eq.~\ref{fgas}, it can be re-written as,

\begin{equation}
\label{gamma}
\gamma(z) = \left[\frac{f_{gas}(z)}{K(z)A(z)}\right]     \left[\frac{\Omega_m(z)}{\Omega_b(z)}\right] \left(\frac{D_A}{D_A^*}\right)^{3/2}.\
\end{equation}
 Simulations  of hot, massive, and dynamically relaxed galaxy clusters ($M_{500}>10^{14} M_{\odot}$) evolving  with  different physical processes have been used to verify the  evolution of $\gamma$ with cluster redshift and  no significant trend of $\gamma(z)$ as a function of redshift has been found \cite{Battaglia2013,Planelles2013}. Recent works have estimated the depletion factor using only cosmological observations, such as: galaxy cluster gas mass fraction (calculated at $r_{2500}$), SNe Ia and strong gravitational lensing (SGL) systems \cite{Holanda2017JCAP,Holanda2018}. In particular, the analysis involving galaxy cluster and SGL systems showed a mild evolution for $\gamma(z)$ \cite{holandadepletion}. On the other hand, constraints on a possible evolution of $\gamma(z)$ using gas mass fraction measurements at $r_{500}$ have also been obtained through a joint analysis with cosmic chronometers \cite{Zheng2019EPJC,boraepjc}. In these works, it was pointed out  that  one cannot use  $f_{gas}$ values  at $r_{500}$ as a stand-alone probe  for any model-independent cosmological tests since these studies have found a non-negligible $\gamma(z)$ evolution.  In this way, before we performed our test on a possible  evolution  of dark matter density law using galaxy clusters (calculated at $r_{2500}$) with cosmic chronometers, it is necessary to verify if these data support  a constant $\gamma$ or point to an  evolving $\gamma$ with redshift. We note that ~\citet{Mantz:2014xba} also looked for a similar redshift variation of $\gamma$ using only the gas mass fraction measurements and found negligible variation.  Here, we do this test in a model-independent method using chronometers, as discussed below.

In order to derive the angular diameter distance to each galaxy cluster, we use 31 cosmic chronometer $H(z)$ data from \citet{li19} in the redshift range $0.07 \leqslant z \leqslant 1.965 $. For this purpose, we choose   Gaussian Processes Regression~\cite{Seikel} to reconstruct the angular diameter distance at each cluster's redshift(for more details, see~\cite{HS,boraepjc,boracddr}). The reconstructed angular diameter distance is obtained via,

\begin{equation}
D_{A} (z) =  \left(\frac{c}{1+z}\right)\int_{0}^{z}\frac{dz^{'}}{H(z')},
\label{eq:daz}
\end{equation}
where $H(z')$ is the non-parametric reconstruction of Hubble parameter using Gaussian Processes.

\begin{figure}[t]
    \centering
    \includegraphics[width=10cm,height=8cm]{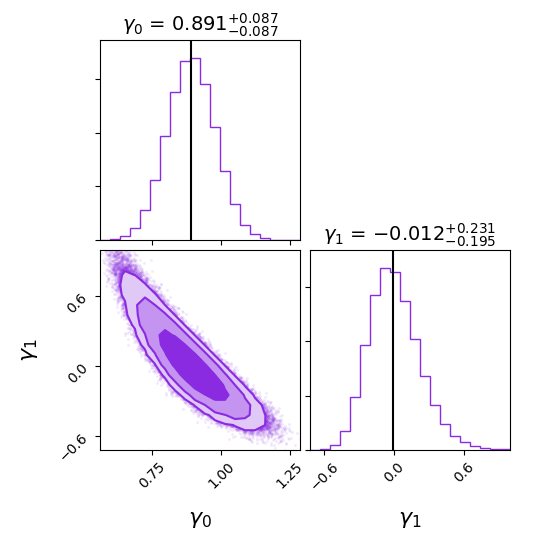} 
    \caption{The 1-D marginalized likelihood distributions along with 2-D marginalized constraints showing the 68\%, 95\%, and 99\% confidence regions for the parameters $\gamma_0$ and $\gamma_1$, obtained using the {\tt Corner} python module~\cite{corner}.}
    \label{fig:fig1}
\end{figure}

\begin{figure}[t]
    \centering
    \includegraphics[scale=0.8]{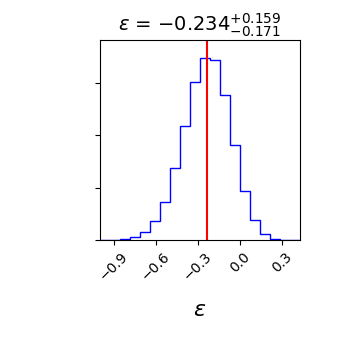} 
    \caption{The likelihood distribution for  $\epsilon$ obtained after maximizing the likelihood in Eq.~\ref{eq:logL2}.}
    \label{fig:fig2}
\end{figure}

\begin{figure*}[t]
    \centering
    \includegraphics[width=12cm, height=8cm]{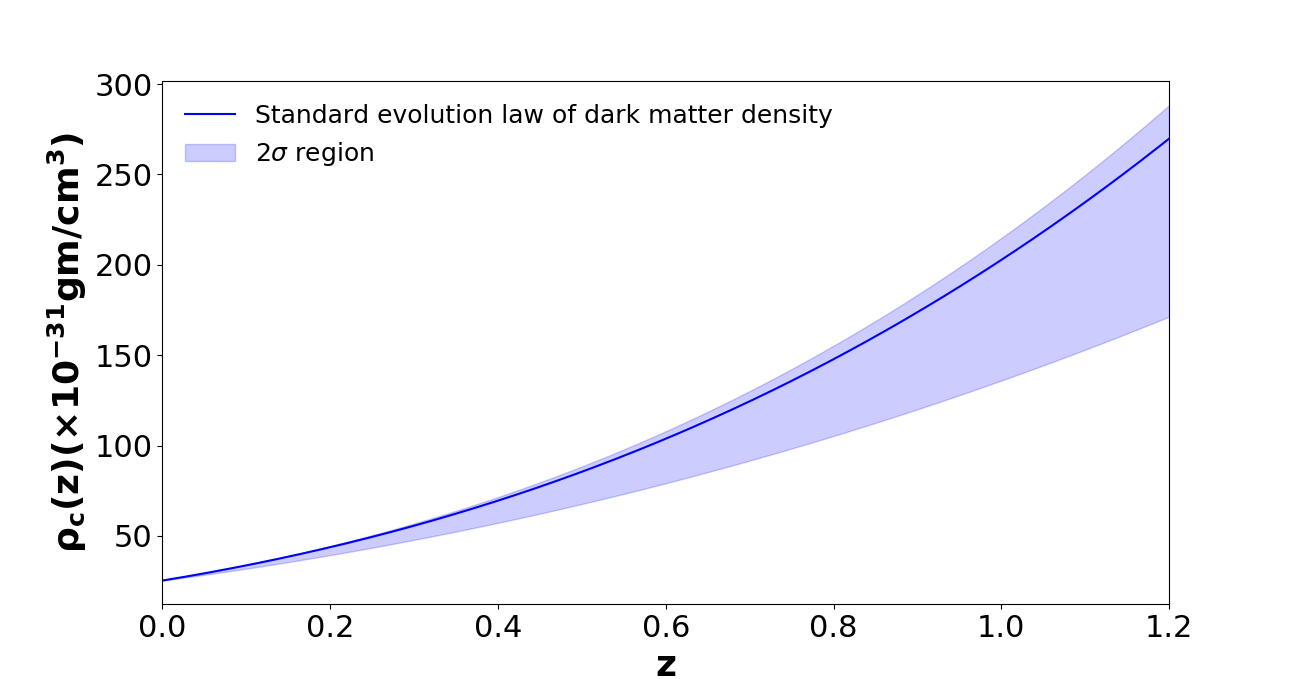} 
    \caption{The evolution law for the dark matter density as a function of redshift. The light blue shaded area shows the  $2\sigma$ allowed region for the evolution law as found in this work. The standard evolution law for the dark matter density is indicated by the solid blue line.}
    \label{fig:fig3}
\end{figure*}

Now, we adopt the linear parametric form to model the gas depletion factor $\gamma(z)$ (which is widely used in our previous works \cite{boraepjc,holandadepletion}) as follows,  

\begin{equation}
\label{gamma_parametric}
\gamma(z) = \gamma_0(1+\gamma_1 z),
\end{equation}
where $\gamma_0$ is the normalization constant and $\gamma_1$ shows the redshift dependent part of the gas depletion factor.

\subsection{Testing a possible evolution  of dark matter density law}

The total matter density of the universe is the summation of baryonic and dark matter density, and  so Eq.~\ref{fgas} can be rewritten as,  
\begin{equation}
\label{fgas1}
f_{gas}(z) = K(z)A(z)\gamma(z)  \left[\frac{\Omega_b(z)}{\Omega_b(z) + \Omega_c(z)}\right] \left(\frac{D_A^*}{D_A}\right)^{3/2}.\
\end{equation}
If we posit  putative dark matter interactions with the remaining dark sectors, then an ad-hoc term $\epsilon$ can be added to the standard dark matter density evolution law, and the deformed evolution law can then be written as: $\rho_c = \rho_{c0} (1+z)^{3+\epsilon}$ where $\epsilon$ represents the interaction term arises due to DM-DE interactions~\cite{wang05,alcaniz05,boraDM}. So Eq.~\ref{fgas1} can be recast as, 
\begin{equation}
\label{epsilon}
(1+z)^{\epsilon} = \left[\frac{\rho_{b0}}{\rho_{c0}}\right] \left[\left(\frac{K(z)A(z)\gamma(z)}{f_{gas}(z)}\right)   \left(\frac{D_A^*}{D_A}\right)^{3/2} - 1 \right],
\end{equation}
where $\rho_{b0}$ and $\rho_{c0}$ are the baryonic matter density and dark matter density respectively at the current epoch. This is the key equation that will be used to constrain the $\epsilon$ parameter.

\begin{table*}[t]
\caption{\label{tab:table1}. Constraints on the parameter $\epsilon$ used in Eq.~\ref{epsilon} as discussed in Sec~\ref{methodology} along with the recent studies.}
    \centering
    \begin{tabular}{|l|c|c|c|r|} \hline
    \textbf{Dataset Used} & \boldmath$\epsilon$ & \textbf{Reference}\\ \hline 
      Gas Mass Fraction + SNe Ia &  $0.13\pm0.235$ & \citet{Holanda:2019sod}\\
       Gas Mass Fraction + Strong Gravitational Lensing & $-0.088\pm 0.11$ & \citet{boraDM}\\
       Gas Mass Fraction + Cosmic Chronometers & $\mathbf{-0.234^{+0.159}_{-0.171}}$ & \textbf{This work}\\

      \hline 
      
    \end{tabular}

\end{table*}

\section{Analysis and Results} 
\label{sec:analysis}

\subsection{Constraints on the  gas depletion factor}

To constrain the parameters $\gamma_0$ and $\gamma_1$ of Eq.~\ref{gamma_parametric}, we need to maximize the likelihood equation which is given as follows,

\begin{widetext}
\begin{equation}
    \label{eq:logL1}
   -2\ln\mathcal{L} =  \sum_{i=1}^{n}\frac{\left( \gamma_0(1+\gamma_1z) - \left[\frac{f_{gas}(z)}{K(z)A(z)}\right]     \left[\frac{\Omega_m(z)}{\Omega_b(z)}\right] \left(\frac{D_A}{D_A^*}\right)^{3/2}\right)^2}{\sigma_{i}^2} + \sum_{i=1}^{n} \ln 2\pi{\sigma_{i}^2}. 
\end{equation}    
\end{widetext} 
Here, $\sigma_i^2$ includes the error in $f_{gas}$, $K(z)$, $\Omega_m(z)$, $\Omega_b(z)$, $D_A$ and $D_A^*$ which is calculated by error propagation of aforementioned quantities.

\subsection{Constraints on $\epsilon$ parameter}

Similarly, the constraints on the $\epsilon$ parameter present in Eq.~\ref{epsilon} can be obtained by maximizing the likelihood distribution function, ${\cal{L}}$  given by

\begin{widetext}
\begin{equation}
    \label{eq:logL2}
   -2\ln\mathcal{L} =  \sum_{i=1}^{n}\frac{\left((1+z)^{\epsilon} - \left[\frac{\rho_{b0}}{\rho_{c0}}\right] \left[\left(\frac{K(z)A(z)\gamma(z)}{f_{gas}(z)}\right)   \left(\frac{D_A^*}{D_A}\right)^{3/2} - 1 \right]\right)^2}{\sigma_{i}^2} + \sum_{i=1}^{n} \ln 2\pi{\sigma_{i}^2}. 
\end{equation}  
\end{widetext} 
Here, $\sigma_i^2$ denotes the observational errors which is obtained by propagating the errors in $\rho_{b0}$, $\rho_{c0}$, $K(z)$, $\gamma(z)$, $f_{gas}(z)$, $D_A^*$ and $D_A$.

\subsection{Results}

The next step is to estimate the free parameters viz. $\gamma_0$ and $\gamma_1$ by maximizing the likelihood function (Eq.~\ref{eq:logL1}) using the $\tt{emcee}$ MCMC sampler~\cite{emcee}. The outcome of this maximization is shown in Fig.~\ref{fig:fig1}. The diagonal histograms represent the one dimensional marginalized posterior distributions for each parameter and the off-diagonal entries show  the 68\%, 95\%, and 99\%  two-dimensional  marginalized confidence intervals. We find that  $\gamma_0 = 0.891\pm0.087$ and $\gamma_1 = -0.012_{-0.195}^{+0.231}$. Our result imply no evolution for the gas depletion factor $\gamma(z)$ as a function of redshift  within $1\sigma$ at $R_{2500}$.

Similarly, we maximize the likelihood function (Eq.~\ref{eq:logL2}) to estimate the free parameter $\epsilon$ present in Eq.~\ref{epsilon}. Since our resulting  $\gamma_1$ from our first analysis  is consistent with zero at $1\sigma$,  we set $\gamma(z) = \gamma_0$ in Eq.~\ref{eq:logL2}. The allowed region for $\epsilon$ is shown in Fig.~\ref{fig:fig2}. We obtain: $\epsilon = -0.234^{+0.159}_{-0.171}$ which shows no interaction of dark matter with the dark sectors to within $1.4\sigma$. The evolution of the dark matter density $\rho_c(z)$ as a function of redshift along with the standard dark matter density evolution law ($\epsilon = 0$) is displayed in Fig.~\ref{fig:fig3}. Table~\ref{tab:table1} summarizes our results and other recent estimates on $\epsilon$ presented in literature.

\section{Conclusions}
\label{sec:conclusions}

The aim of this work is twofold: First, we studied a possible evolution of the gas depletion factor using 35 X-ray gas mass fraction data from \citet{Mantz:2014xba} in the redshift range $0.078 \leq z \leq 1.063$ at near $R_{2500}$, in a model-independent manner. The key equation used for this analysis is Eq.~\ref{gamma}. In order to derive the angular diameter distance to each cluster, we incorporated the cosmic chronometer data from \cite{li19} spanning the redshift range $0.07 \leqslant z \leqslant 1.965 $ and then applied Gaussian Processes to estimate the angular diameter distance for each cluster redshift, as given by Eq.~\ref{eq:daz}. Furthermore, we adopted a linear parametric form for the gas depletion factor, $\gamma(z) = \gamma_0(1+\gamma_1 z)$. We constrained the $\gamma_0$ and $\gamma_1$ parameters by maximizing the log-likelihood given by Eq.~\ref{eq:logL1}. The one dimensional marginalized likelihoods for  each parameter along with two dimensional confidence intervals are shown in Fig.\ref{fig:fig1}. We found no evolution of the gas depletion factor as a function of redshift within $1\sigma$ at near $R_{2500}$.

In the second part, we performed a test to study a possible departure of the standard dark matter density evolution law, $\rho_c = \rho_{c0} (1+z)^3$, as a followup to our previous studies which did this test using combination of gas mass fraction measurements and Type 1a supernova~\cite{Holanda:2019sod} as well as strong lensing systems~\cite{boraDM}.
For this purpose, we added an ad-hoc term $\epsilon$ to the evolution law: $\rho_c = \rho_{c0} (1+z)^{3+\epsilon}$ where $\epsilon$ represents the interaction term arises due to DM-DE interactions. To constrain the $\epsilon$ parameter presented in Eq.~\ref{epsilon}, we maximized the log-likelihood  in Eq.~\ref{eq:logL2}. Note that we used  a constant value of $\gamma(z)$,  obtained from the first analysis of this work. Fig.~\ref{fig:fig2} shows the one dimensional likelihood for $\epsilon$ parameter. Our final result from this analysis is: $\epsilon = -0.234^{+0.159}_{-0.171}$ indicating no interaction of DM-DE at $1.4\sigma$. This is shown in Fig.~\ref{fig:fig3} along with the standard DM evolution law($\epsilon=0$) as a function of redshift. Constraints on the $\epsilon$ obtained in recent studies along with our result are summarized in Table~\ref{tab:table1}. 
The results in this work are in agreement with those from previous analyses.

\section*{ACKNOWLEDGEMENT}
KB acknowledges the Department of Science and Technology, Government of India for providing the financial support under DST-INSPIRE Fellowship program. RFLH
thanks CNPq No.428755/2018-6 and 305930/2017-6. SHP would like to thank CNPq - Conselho Nacional de Desenvolvimento Cient\'ifico e Tecnol\'ogico, Brazilian research agency, for financial support, grants number 303583/2018-5.

\bibliography{ref}

\end{document}